\documentstyle[preprint,aps]{revtex}
\tighten
\def\go{
\mathrel{\raise.3ex\hbox{$>$}\mkern-14mu\lower0.6ex\hbox{$\sim$}}
}
\def\lo{
\mathrel{\raise.3ex\hbox{$<$}\mkern-14mu\lower0.6ex\hbox{$\sim$}}
}
\begin{document}
\title{Knots in Magnetized Relativistic Jets and
Their Simulations\footnote{SUBMITTED TO NATURE}}
\author{Maurice H.P.M. van Putten\footnote
	{Electronic address: \tt vanputte@spacenet.tn.cornell.edu\hfil}}
\address{CRSR \& Cornell Theory Center\\
        Cornell University\\
	Ithaca, NY 14853-6801}
\date{\today}
\maketitle
\baselineskip24pt
\mbox{}\\
{\bf Bright features (`knots') in extragalactic jets
are commonly associated with reheating of charged
particles in regions with shocks or strong compression.
We find a natural appearance of
nozzles in simulations of toroidally magnetized jets,
and propose this as a generic mechanism
of knotted structures such as in the jet of 0800+608.
Nozzles provide correlated
enhancement of emission in radio and optical,
such as discovered in HST observations of 3C273.
The simulations reveal a rich structure of knots, including
a nose cone with a propagating knotted structure
and a hot-spot as the most extended brightness feature.
Unexpected is the observed separation of the
hot-spot from a composite Mach disk closer to the source.}

Enhanced brightness features in the form of localized
`knots' are perhaps the most significant morphological
aspects of astrophysical jets. These features are produced
with relativistic motion in both
extragalactic sources
($e.g.$ 3C273$^1$ and M87$^2$)
and recently discovered galactic sources
(GRO J1655-40$^{3,4}$ and GRS 1915-105$^{4,5}$).
On the larger scale,
the most extended brightness feature
(the `hot-spot')
has been explained as enhanced synchrotron radiation
behind a terminal shock$^{6,7}$.
Increased resolution and sensitivity of radio observations shows
bright features over the entire length of some jets,
from close to the source
up to the terminal hot-spot, which
can be interpreted likewise$^{8,9,10,11}$.
Knotted structures appear with extraordinary dynamic range
in brightness and patterns, including
knots as the first detectable features
at finite angular distance away from the source
(leaving an apparent gap between source and jet,
as $e.g.$ in M84$^{12}$),
and regular or quasi-periodic knotted structures
which characterize the large scale morphology
($e.g.$ as in 3C 111$^{13}$, QSO 0800+608$^{14,15}$
(see Figure 1),
Cyg A$^{16}$).
If knot motion indeed relates to bulk motion,
BL Lacertae objects, radio quasars and radio galaxies
emit plasma with appreciable Lorentz
factors$^{17}$.
Relativistic effects introduce beaming
($cf.$ Ref 11),
which allows for the unification scheme for compact
radio loud sources.

In this $Letter$, the nature of knots is considered in the large scale
morphology of jets
modeled by the theory of time-dependent, fully relativistic
ideal magneto-hydrodynamics (MHD) of a polytropic, perfect fluid
as the result of boundary conditions associated with
the `standard model' of active galactic nuclei
(a rotating black hole and accretion disk$^{24,25}$;
see also Refs 26 and 21 for a discussion).
The simulations use the equations of
MHD in divergence form$^{18-23}$.

Quasi-periodic structures have a natural appearance in
simulations of toroidally magnetized relativistic jets,
which are shed off by an oscillation in the hot-spot
in the head of the jet subsequent to
a stagnation point near the source (Figure 2$^{21}$).
Figure 2 shows a stagnation point (first, left pressure maximum),
and three subsequent local pressure maxima with Mach number
$M=1$, which are $sonic$ $nozzles$ with
pressure and density contrasts
$4.39^{\pm1.37}$ and $2.67^{\pm0.57}$, respectively.
The nozzles constitute
funnels of nearly
unmagnetized flow, pinched by the
magnetized flow at larger radius; since the toroidal magnetic field vanishes
on axis, it is dynamically appreciable only some distance away from
the axis and the flow near-axis is to leading order purely hydrodynamic.
The basic scaling of the nozzles can be derived by
considering a tube of streamlines through a characteristic cross section,
$A_s$, at the sonic point,
and its (for supersonic flow wider) cross section, $A$,
in between two nozzles.
The density contrast $\frac{r}{r_s}$ as a function of the
ratio of widths
$\alpha=\sqrt{\frac{A}{A_s}}$ is parametrized
by the sound speed, $a_s$, at the sonic point.
The nonrelativistic ($a_s\rightarrow0$)
and ultra-relativistic ($a_s\rightarrow\sqrt{\gamma-1}$)
limits thus provide upper and lower bounds, respectively, for $\frac{r}{r_s}$
close to the axis,
which for large $\alpha$ are asymptotically given by
\begin{eqnarray}
\alpha^{-\frac{2}{2-\gamma}}
(\gamma-1)^{\frac{1}{2(2-\gamma)}}
\lo\frac{r}{r_s}\lo\alpha^{-2}\sqrt{\frac{\gamma-1}{\gamma+1}}.
\end{eqnarray}

Figure 3 shows the result of a supersonic ($M=1.67)$, low density
%toroidally magnetized relativistic
relativistic ($\Gamma=2.46)$ jet with out of radial force-balance
boundary conditions due to a toroidal magnetic field
as the ``spin-off" in the energy
extraction process by poloidal field lines in the source.
The jet propagates into an unmagnetized outer
medium at rest and in hydrostatic pressure balance
with the jet.
The central jet flow partially thermalizes in a Mach disk,
which consists of a shock of small radius,
which is by-passed by jet flow which shocks in an annular shock
further downstream. In between these two shocks resides
a much weaker and larger shock wave,
the combination of which
with the small Mach disk we shall refer to as the composite
Mach disk.
The shocked by-pass flow subsequently bifurcates into a cocoon
and a recollimated nose cone. The cocoon exerts a pinch on
the boundary of the jet in the neighborhood of the
small Mach disk, which initiates the formation of a weakly shocked nozzle;
a feedback
loop is thus obtained (see Ref 27 for a feedback loop
in hydrodynamical jets). Further refinement of the computations
are needed to detail this process.
At the `root' of the nose cone resides a stagnation point
(as in the nonrelativistic simulations
of Refs 28 and 29),
which sets the stage for the formation of nozzles as in
Figure 2 {\em in the nose cone ahead of the Mach disk}.
A hot-spot results
as the most extended enhanced brightness feature, separated off the
composite Mach disk and flow bifurcation structure.
Exploration in parameter space is needed to obtain critical
values at which this morphology is typical.
Because the nozzles form on axis by the internal dynamics of
the jet, the formation process is expected to be rather insensitive
to (the interaction with) the environment,
and hence the process is expected to persist
in three-dimensional simulations ($i.e.$ with no imposition of
cylindrical symmetry).
This simulation illustrates that we can learn from the
extended morphology (up to the terminal hot-spot) about the
characteristics of the
flow at the source (the Mach number and
radial distribution of toroidal magnetic field),
which may serve as a new approach towards probing
the energy extraction process
in the core of active galactic nuclei.

If the knots of 0800+608 turn out to be (i)
standing still, our results predict the
jet to be transonic (with a stagnation point
close to the source; Figure 2); (ii) moving, our results
predict the jet to be supersonic (with the
stagnation point moving ahead of the Mach-disk; Figure 3).
The asymmetry in the morphology of
0800+608 has been attributed by Jackson $et$ $al.^{15}$ to a
wind interacting disruptively with the counter jet, which favors
(i) if the jets are symmetric at the source.
In alternative (ii), the recollimation associated with the
stagnation point may denote a transition to
constant width in an environment of decreasing pressure
(knot 3 in Figure 1).
The weak extended structure beyond
knot 8 then corresponds to part of the
thin cocoon around the nose cone in Figure 3.
Jackson $et$ $al.^{15}$ infers the presence of
a cocoon between knot 3 and knot 8 from edge-brightening
due to jet-cocoon interaction, which would need further refinement
in our model to be brought about in emissivity.
Our results support the suggestion of Jackson $et$ $al.^{15}$
of the knots in 0800+608 being associated with
an intrinsic, stable and periodic magnetic pinch, as opposed
to being due to an unstable interaction with the environment.

The adiabatic process associated with the sonic knots
finds a place in the observed correlations
between radio and (HST-)optical synchrotron emission
in the recent observations on 3C273$^{30,31}$.
The relativistic simulations reported here also serve as a step towards
fully three-dimensional simulations
with helical magnetic fields, as required in the unification scheme
in compact radio loud sources.
\newpage
\centerline{\bf REFERENCES}
\mbox{}\\
\mbox{ }1. Pearson T.J., Unwin S.C. Cohen M.H., Linfield R.P., Readhead A.C.S.,
Seielstad G.A.,\\
\mbox{  }\mbox{  }\mbox{  }\mbox{  }Simon, R.S. \&
Walker R.C., {\em Nature}, {\bf 290}:365-368 (1981).\\
%\mbox{ }1. Flatters C. \& Conway R.G., {\em Nature}, {\bf 314}:425 (1985).\\
%Flatters C. \& Conway R.G., {\em Nature}, {\bf 314}:425 (1985).\\
\mbox{ }2. Birretta J.A., Zhou Z. \& Owen F.N., {\em Ap. J.},
{\bf 447}(2):582-596 (1995).\\
\mbox{ }3. Hjelming R.M. \& Rupen M.P,
	{\em Nature}, {\bf 375}(8):465-468 (1995).\\
\mbox{ }4. Mirabel I.F. \& Rodriguez,
	{\em in} VII-th Texas Symposium on relativistic astrophysics
	(1995).\\
\mbox{ }5. Levinson A.\& Blandford R.D.,
	{\em Ap. J. Lett.} (to appear) (1995).\\
\mbox{ }6. Scheuer P.A.G.,
	{\em Mon. Not. R. Astr. Soc.}, {\bf 166}:513-528 (1974).\\
\mbox{ }7. Blandford R.D. \& Rees M.J.,
	{\em Mon. Not. R. Astron. Soc.}, {\bf 169}:395-415 (1974).\\
\mbox{ }8. Rees J.M., {\em Mon. Not. R. Astron. Soc.} {\bf 184}(3):61P-65P
(1978).\\
\mbox{ }9. Blandford R.D. \& Konigl A.,
	{\em Ap. J.}, {\bf 232}:34-48 (1979).\\
10. Cawthorne T.V., Wardle J.F.C., Roberts D.H. \& Gabuzda D.C.,
        {\em Ap. J.} {\bf 416}:519-535,\\
\mbox{   }\mbox{   }\mbox{   }\mbox{   } (1993).\\
11. Krishna G.\& Wiita P.J., {\em Nature}, {\bf 363}:142-144 (1993).\\
12. Bridle A.H.\& Perley R.A.,{\em Ann. Rev. Astron. Astrophys.},
{\bf 22}:319-358(1984).\\
13. Linfield R. \& Perley R.A., {\em Ap. J.}, {\bf 279}:60-73 (1984).\\
14. Shone D.L. \& Browne I.W.A., {\em Mon. Not. R. Astron. Soc.},
        {\bf 222}:365-372 (1986).\\
15. Jackson N., Browne I.W.A., Shone D.L. \& Lind K.R.,
        {\em Mon.Not.R.Astron.Soc.}, {\bf 244}:
\mbox{   }\mbox{   }\mbox{   }\mbox{   }750-758 (1990).\\
16. Perley R.A., Dreher J.W. \& Cowan J.J., {\em Ap. J. Lett.}, {\bf
285}:L35-L38 (1984).\\
17. Ghissilini G., Padovani P., Celoti A \& Maraschi,
	{\em Ap. J.}, {\bf 407}:65-82 (1993).\\
18. van Putten M.H.P.M.,
	{\em Commun. Math. Phys}, {\bf 141}:63-77 (1991);\\
19. \mbox{ }\mbox{ }--,{\em J. Comput. Phys.}, {\bf 105}(2):339-353 (1993);\\
20. \mbox{ }\mbox{ }--,{\em Ap. J. Lett.}, {\bf 408}:L21-L23 (1993);\\
21. \mbox{ }\mbox{ }--,$in$ Proc.~C.Lanczos~Int.~Centenary~Conf.,
        Moody~C.,Plemmons~R.,Brown~D.~and~\\
\mbox{  }\mbox{  }\mbox{  }\mbox{  } Ellison D. (eds.), p449-451,
SIAM (1994);\\
22. \mbox{ }\mbox{ }--,{\em Internat. J. Bifur. \& Chaos}, {\bf 4}(1):57-69
(1994);\\
23. \mbox{ }\mbox{ }--,{\em SIAM J. Numer. Anal.}, {\bf 32}(5):1504-1518
(1995).\\
24. Lovelace R.V.E., {\em Nature}, {\bf 262}:649-652 (1976).\\
25. Blandford R.D. \& Znajek R.L.,
	{\em Mon. Not. R. Astron. Soc.}, {\bf 179}:433-456 (1977).\\
26. Bellan P.M.,{\em Phys. Rev. Lett.}, {\bf 69}(24):3515-3518 (1992).\\
27. Norman M.L., Smarr L., Winkler K.H.A. \& Smith M.D.,
	{\em Astron. Astrophys.}, {\bf 113}:\\
\mbox{   }\mbox{   }\mbox{   }\mbox{   }285-302 (1982).\\
28. Clarke D.A., Norman M.L. \& Burns J.O., {\em Ap. J. Lett.},
	{\bf 311}:L63-L67 (1986).\\
29. Lind K.R., Payne D.G., Meier D.L. \& Blandford R.D.,
	{\em Ap. J.}, {\bf 344}:89-103 (1989).\\
30. Thomson R.C., Mackay C.D. \& Wright A.E., {\em Nature},
        {\bf 365}:133-135 (1993).\\
31. Bahcall J.N., Kirhakos S., Schneider D.P.,
	Davis R.J., Muxlow T.W.B., Garrington\\
\mbox{  }\mbox{   }\mbox{   }\mbox{  }\mbox{   }S.T., Conway R.G. \& Unwin
S.C.,
	{\em Ap. J. Lett.}, {\bf 452}(2):L91-L94, (1995).\\
\mbox{}\\
ACKNOWLEDGEMENTS.
The author greatfully acknowledges
stimulating discussions with
Robert Antonucci, Amir Levinson, Richard
Lovelace, David Chernoff and Saul Teukolsky.
This
work is supported by grants from NSF and NASA.
The Cornell Theory Center is supported by
NSF, NY State,
ARPA, NIH, IBM and others.
\newpage
\mbox{}\\
{\bf FIGURE 1.}
A 6cm (0.35-arcsec circular beam) total
intensity VLA map of 0800+608
($z$=0.689, $L$=140kpc with H=50kms$^{-1}$Mpc$^{-1}$)
by Jackson $et.$ $al.$ (1990). Contours are from
0.05-6.4 mJy beam$^{-1}$ (Courtesy
of Blackwell Science Ltd).\\
\mbox{}\\
\mbox{}\\
{\bf FIGURE 2.}
%\begin{figure}
%\caption {\small
Coordinate distributions of pressure (A) and magnetic field
strength (B) of a fully developed axisymmetric
relativistic jet
($v\lo 0.73c$, where $c$ is the velocity of light),
produced by a Newtonian sphere
with boundary conditions
$(H,P,r)=(0.72\sin\theta,0.100,0.200)$
(in spherical coordinates with $\theta=0$ on the
axis of symmetry).
The fluid has $\gamma=\frac{3}{2}$, and is
released from the surface of the sphere by diffusion.
The Mach number
$M=\frac{v}{a}$ on-axis (C) (sound speed
$a=\{\frac{\gamma P}{rf}\}^{\frac{1}{2}}$) shows
the formation of sonic nozzles; solid and dashed lines
correspond to two consecutive
times, $15\%$ apart, with local Mach numbers at the
three sonic points (in between brackets give values at the
later time).
The jet shows a stagnation point (first, left knot),
and subsequently three sonic ($M=1$) nozzles
in various degrees of maturation, shed off
by an oscillation in the hot-spot in the head of the jet.
Schematic diagram (D) of a tube of streamlines
in a sonic ($M=1$) nozzle
with cross section $A_s$ at
the sonic point and $A$ at maximal expansion.
The flow is described by
Bernoulli's equation, $H=f^*\Gamma=f_s^*\Gamma_s$, and continuity,
$\Phi=ru^zA=r_ru^zA_s$, where $f^*=f+kr^2$ and the subscript $s$ refers to
the sonic section. Because $k\equiv\frac{h}{r}=O(\sigma)$, the near-axis
flow is to leading order a hydrodynamical flow,
and $f^*\sim f$ facilitates the relationship
of the density contrast $\frac{r}{r_s}$ as a function
of $\alpha=(\frac{A}{A_s})^{\frac{1}{2}}$.
Using $u^b\sim(\cosh\lambda,\sinh\lambda,0,0)$ near axis,
the nonrelativistic limit $a_s\rightarrow0$ gives
$\alpha_0(M)=(\frac{2}{\gamma+1})^{\frac{\gamma+1}{4(\gamma-1)}}
M^{-\frac{1}{2}}(1+\frac{\gamma-1}{2}
M^2)^{\frac{\gamma+1}{4(\gamma-1)}}
\sim(\frac{\gamma-1}{\gamma+1})^{\frac{1}{4}\frac{\gamma+1}{\gamma-1}}
M^{\frac{1}{\gamma-1}}$ as $M\rightarrow\infty$,
where $M=(\frac{r}{r_s})^{-\frac{\gamma+1}{2}}\frac{1}{\alpha^2}$,
which serves as an upper bound on $\frac{r}{r_s}$.
The relativistic case retains the sound speed, $\tanh\lambda_s=a_s$,
at the sonic point as a parameter in
$\alpha_1(\frac{r}{r_s};a_s)=
 (\frac{r}{r_s})^{-\frac{1}{2}}
 \{\frac{f_s^2}{f^2}\cosh^2\lambda_s-1\}^{-\frac{1}{4}}
 \sinh^\frac{1}{2}\lambda_s
 \sim r^{-\frac{(2-\gamma)}{2}}(\gamma-1)^{\frac{1}{4}}$
as $a_s\rightarrow(\gamma-1)^\frac{1}{2}$,
where $f=1+\frac{f_sa^2_s}{\gamma-1}(\frac{r}{r_s})^{\gamma-1}$
with $f_s=(1-\frac{a^2_s}{\gamma-1})^{-1}$.
The computations are in cylindrical coordinates
$(t,\sigma,\phi,z)$,
using the covariant and constraint-free
divergence formulation
of relativistic MHD
(van Putten 1991-1995), and are performed on the
Intel Paragon parallel computer at SDSC, San Diego.
\mbox{}\\
\mbox{}\\
{\bf FIGURE 3.}
Coordinate distributions of velocity magnitude,
pressure, rest mass density and magnetic field strength of
fully developed axisymmetric, toroidally magnetized
jet at time $t/\sigma_{jet}=34.47$, where $\sigma_{jet}$ is
the radius of the jet (in arbitrary units).
The colors vary linearly from blue
to red. The jet aperture has
boundary conditions
$(\Gamma,M,H,P,r)
=(2.46,1.67,0.46\sigma\cos\frac{\sigma}
{\sigma_{jet}}\frac{\pi}{2},0.100,0.200)$
which are out of radial
force-balance, and the environment is unmagnetized
and initially at rest with
$(P,r)=(0.100,1.00)$. The aperture boundary conditions
at $\sigma=\sigma_{jet}$ are smoothed
through multiplication by a tangent hyperbolic
(this incidentally gives a thin layer of
reversed toroidal magnetic field).
The on-axis distributions of pressure
and rest-mass density of the jet vary by a factor of 190
and 45, respectively.
The solution is characterized by a hot-spot
of postshock material
at the head of the nose cone
(as in Refs 6 and 7), which has
separated off from
the composite Mach disk. The oscillation in the terminal shock
produces a propagating
($v=0.162^{\pm0.04}c$), supersonic
nozzle ($M=1.28^{\pm0.03}$ in the co-moving frame)
with pressure contrast 6.14 and rest mass density contrast
3.38.
A repeat of the oscillation in the terminal shock
is observed at $t/\sigma_{jet}=39.06$, which produces a second
nozzle (not shown).
The computations use the same method of that for Figure 2,
and are performed on the
IBM SP2 parallel computer at the Cornell Theory Center.
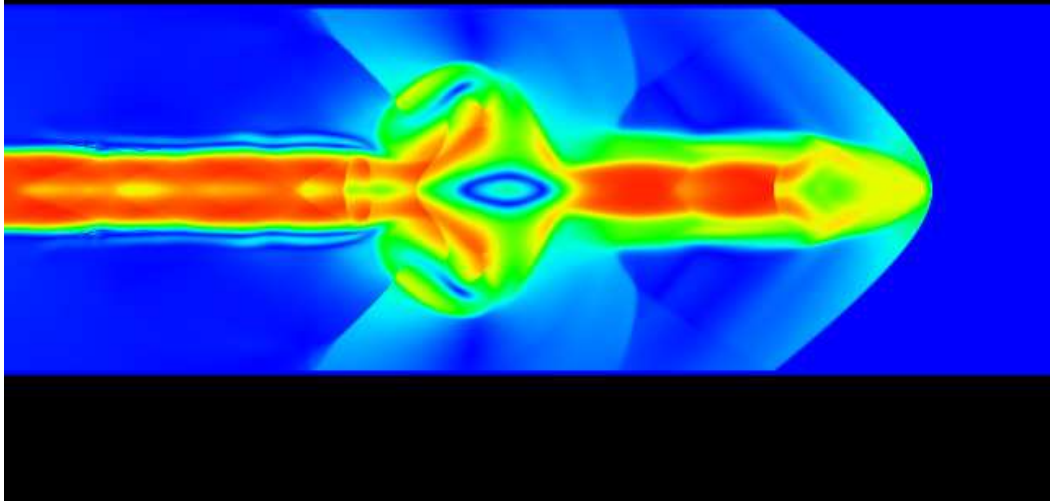
\end{document}